\documentclass[12pt]{article}
\usepackage[english]{babel}

\pdfoutput=1
\usepackage{graphicx}
\usepackage{xcolor}

\usepackage{amssymb,graphicx}
\usepackage{amsmath,amsfonts}
\usepackage{fullpage}

\def\l{\left(}									
\def\r{\right)}

\newcommand{\be}{\begin{equation}}
\newcommand{\ee}{\end{equation}}
\newcommand{\ba}{\begin{align}}
\newcommand{\ea}{\end{align}}
\newcommand{\bg}{\begin{gather}}
\newcommand{\eg}{\end{gather}}
\newcommand{\bseq}{\begin{subequations}}
\newcommand{\eseq}{\end{subequations}}

\newcommand{\Res}{\mathop{\rm Res}\nolimits}

\newcommand{\vk}{{\vec k}}

\title{Unitarity of Minkowski non-local theories made explicit}

\author{Alexey S. Koshelev$^{1}$ and Anna Tokareva$^{2}$
\\ \mbox{}$^{1}$\textit{\small Departamento de F\'isica, Centro de Matem\'atica e Aplica\c{c}\~oes (CMA-UBI),}
\\\textit{\small Universidade da Beira Interior, 6200 Covilh\~a, Portugal}
\\ \mbox{}$^{2}$\textit{\small Department of Physics, University of Jyv\"askyl\"a, P.O.Box 35 (YFL), FIN-40014, Finland {\rm \&}}
\\\textit{\small Institute for Nuclear Research of Russian Academy of Sciences, 117312 Moscow, Russia {\rm \&}}
\\\textit{\small Helsinki Institute of Physics (HIP), University of Helsinki,
P.O. Box 64, 00014, Finland}
}

\date{~}

\begin{document}
\maketitle

\begin{abstract}
  In this work we explicitly show that the perturbative unitarity of analytic infinite derivative (AID) scalar field theories can be achieved using a modified prescription for computing scattering amplitudes. The crux of the new prescription is the analytic continuation of a result obtained in the Euclidean signature to the Minkowski external momenta. We intensively elaborate an example of a non-local $\phi^4$ model for various infinite derivative operators.
 General UV properties of amplitudes in non-local theories are discussed.
\end{abstract}

\section{Introduction}

Higher-Derivative theories appear in different contexts \cite{Ostro:1850,Efimov:1967pjn,Coleman:1968wh,Stelle:1976gc,Fradkin:1981hx,Kuzmin:1989sp,Krasnikov:1987yj,Tomboulis:1997gg,Brekke:1988dg,Vladimirov:1994wi,Dragovich:2017kge,Anselmi:2018kgz,Nicolis:2008in,Horndeski:1974wa,Kobayashi:2019hrl,Biswas:2011ar,Biswas:2016egy,Keltner:2015xda,Koshelev:2020xby,Koshelev:2020fok,Boos:2019vcz} from pure phenomenological models to fundamental theories with the notable example of string field theory \cite{Witten:1985cc,Witten:1986qs,Ohmori:2001am,Arefeva:2001ps} which belongs to the particular interesting class of models featuring analytic infinite derivative (AID) operators entering Lagrangians and acting as form factors. Analyticity of these operators at low momenta guarantees a smooth local IR limit and prompts for the UV modification of respective theories. 

A model example providing such a higher derivative modification can be readily written as follows:
\be 
\label{L}
S=\int d^4x \left[{-\frac{1}{2}\phi (\square-m^2)f(\Box)^{-1}\phi-\frac{\lambda}{4!}\phi^4}\right]
\ee
where $\square$ is the d'Alembertian operator.
This action is the subject of our consideration in this paper which is aimed at illustrating explicitly that the unitarity can be maintained in this class of models.
The chosen model represents perhaps the simplest model featuring the properties which are ought to be studied and understood. Namely the propagator is manifestly of an infinite order in derivatives and there is an interaction term which generates a bounded from below potential. These are the features typical for non-local models that have applications to quantum gravity (renormalizable and ghost-free AID theories) \cite{Koshelev:2020xby} and UV finite non-local scalar theories with an arbitrary potential \cite{Pius:2016jsl,Koshelev:2020fok}.
It is important and interesting to perform similar computations in models more closely related to the AID gravity theories, i.e. for other spins and more general interactions but this is beyond the scope of the current study.

Having the metric signature fixed as $(+---)$ the above Lagrangian describes a normal non-ghost field provided $f=1$ which is also not a tachyon as long as $m^2>0$. Presence of extra derivatives may generically spoil the model by the appearance of ghosts in the spectrum. A simple way to avoid this is to demand that $f(\square)^{-1}$ is an exponent of an entire function. Then the system has no new degrees of freedom as there are no finite poles of the propagator apart from the already existing one at the point $m^2$. This picture seems to be good especially given that higher-derivative factor can easily provide better convergence of loop integrals.

Here exactly a serious issue arises because an exponent of an entire function must have an essential singularity at complex infinity making the use of the Wick rotation unjustified \cite{Peskin:1995ev}. Indeed, the latter requires that the propagator has a pole at the complex infinity, not an essential singularity and this cannot be achieved in our model.

The situation, however, has a resolution presented in the paper by Pius and Sen \cite{Pius:2016jsl} for the class of AID theories originating from the string field theory. The resolution consists of certain modifications to the prescription of how the loop integrals are computed. Namely specially designed integration contours for integrals over the loop momenta are prescribed.
It is interesting to note that related idea was presented many years ago by Efimov \cite{Efimov:1967pjn}.
To be phrased in short, the idea of Efimov is to perform all the inner loop integrals assuming the Euclidean signature of all the momenta. After a result is obtained one should continue analytically this result for the computed amplitude to Minkowski signature for all external momenta. This procedure was claimed to gain unitary scattering amplitudes.

In the present work, using the simplest example of the fish diagram in $\phi^4$ theory we illustrate that the idea by Efimov in fact leads to the same definition of the amplitude as the method elaborated by Pius and Sen\footnote{See also \cite{Pius:2018crk,Briscese:2018oyx}.}. The latter provides rules how to account for the poles at the complex plane when integrating over loop momenta. These rules are actually the same as in local theories, since non-local factors do not lead to new poles. 
Let us also mention that the definition of amplitudes by formal application of the familiar Wick rotation (which is the same as doing the analytic continuation of the result upon completion of loop computations) yields a well-defined local limit. In this limit, the normal Wick rotation is recovered, as well as the unitarity of a local theory.

In our analysis we first provide an explicit example of computations fixing $f(\square)=\exp(\alpha\Box)$ which already comprise a highly non-trivial set of formulae and then move on by showing that one can integrate out internal momenta at one-loop level for an arbitrary form-factor $f(\Box)$. This leads us to an extensive discussion on UV properties of the amplitudes in non-local theories. We then end up by conclusion and outline.

%%%%%%%%%%%%%%%%%%%%%%%%%%%%%%%%%%%%%%%%%%%%%%%%%%%%%%%%%%%%%%%%%%%%
\section{Fate of the Wick rotation}

In a local quantum field theory the amplitudes can be defined in Minkovski space-time within
the familiar rule for the poles. Than it is convenient to do a Wick rotation and go to Euclidean
momenta. In non-local theory, however, it is often problematic to define the Minkowski
amplitude because the corresponding integrals may diverge. In this Section we show that there
is no feasible way to choose a special non-local form-factor which would allow for obtaining convergent Minkowski amplitudes.

\begin{figure}[htb]
    \begin{center}
  \includegraphics[scale=1]{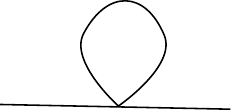}
\caption{Tadpole one-loop contribution to the propagator in $\phi^4$ theory.}
\label{fig:tadpole}
\end{center}
\end{figure}

Let us start with the simplest tadpole correction to the propagator depicted in Fig.~\ref{fig:tadpole}.

Formally, in Minkowski space we can
write
\be 
{\cal A}=\int{\frac{d^4 k f(k_0^2-\vk^2)}{k_0^2-\vk^2-m^2}}
\ee
Than, we can do the Wick rotation and obtain the Euclidean integral,
\be 
{\cal A}_E=-i\int{\frac{d^4 k_E f(-k_{0E}^2-\vk^2)}{k_{0E}^2+\vk^2+m^2}}
\ee
where $k_{0E}=-ik_0$.
This integral can be made manifestly convergent due to an appropriate choice of the function $f(\Box)$. However, we can not conclude that ${\cal A} = {\cal A}_E$ unless we check that the integral over an infinite arc
\be 
{\cal A_C}=\int_0^{\frac{\pi}{2}}{\frac{d^3 k R e^{i\theta}d\theta f(R^2 e^{2 i \theta}-\vk^2)}{R^2 e^{2 i \theta}-\vk^2-m^2}}
\ee
vanishes in the limit $R\rightarrow \infty$. In this limit,
\be 
\begin{split}
{\cal A^{\infty}_C}&=\int_0^{\frac{\pi}{2}}{\frac{ R e^{i\theta}d\theta f(R^2 e^{2 i \theta})}{R^2 e^{2 i \theta}}}\\
&=\int^R_r{\frac{i f(-z^2)}{z^2}}+\int^R_r{\frac{f(z^2)}{z^2}}+\int^{\frac{\pi}{2}}_0 f(0)\frac{e^{-i\theta}}{r}d\theta\\
&=\int_r^R\frac{f(0)(i+1)-f(z^2)-i f(-z^2)}{z^2}
\end{split}
\ee
If $f(\square)=1$ than this integral is zero which makes the Wick rotation consistent in a local
theory. However, in non-local models the situation is less obvious. If ${\cal A^{\infty}_C}=\mathrm{const}<\infty$ than ${\cal A_C}=\int{\cal A^{\infty}_C}d^3 \vk$ is divergent which means that the Minkowski amplitude is divergent too.
In principle it is possible to choose $f(\Box)$ in such a way that the condition
\be 
\int \frac{(f(0)(i+1)-f(z^2)-if(-z^2))dz}{z^2}=0
\ee
is satisfied. However, when considering other diagrams, we obtain other conditions, finally an infinite set of similar conditions on $f(\square)$ which a priori would not be resolved simultaneously.
We thus conclude, that it is nearly impossible to define non-local amplitudes in Minkowski space in a standard way unless a mysterious combination of a form-factor and the theory potential can be found such that an infinite tower of conditions allowing the Wick rotation is resolved. Independently of the form of the non-local propagator, the familiar Wick rotation can not be done.
Therefore, another way to define physical non-local scattering amplitudes is required.

\section{Unitarity of the fish diagram}
\label{sec:unitarity}

As we have shown just before, the definition of physical amplitudes in non-local theories is a tricky point which most likely cannot be resolved using standard local field theory prescriptions by just making an appropriate choice of the non-local propagator. We thus turn to the method elaborated in papers \cite{Efimov:1967pjn,Pius:2016jsl}.

We are going to study the procedure using the so-called fish diagram in non-local theory \eqref{L}. This diagram depends on external momenta in a combination $p=p_1+p_2$ but choosing the center of mass reference frame we can reduce the number of variables to only one scalar $p_0$. The corresponding Euclidean matrix element would be defined as,
\be 
{\cal M}_E=-\frac{\lambda^2}{32\pi^4}I(p_E)
\ee
The physical matrix element is to be defined as an analytic continuation of ${\cal M}_E(p_E)$ to Minkowski momenta $p$.

    \begin{figure}[htb]
    \begin{center}
  \includegraphics[scale=.8]{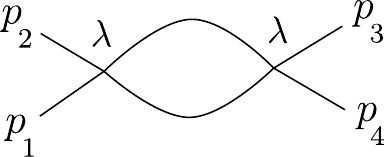}
\caption{One-loop contribution to $2\rightarrow 2 $ scattering in $\phi^4$ theory.Here only $s$-channel contribution is shown, $p=p_1+p_2$.}
\label{fig:fish}
\end{center}
\end{figure}

Here $I(p_E)$ is defined as Euclidean integral
\be 
I(p_E)=\int d^4 k_E \frac{f(k_E^2)\,f((p_E-k_E)^2)}{(k_E^2+m^2)((p_E-k_E)^2+m^2)}
\ee
The integral can be rewritten as
\be 
I(p_E)=2\int\frac{f(k_E^2)\,f((p_E-k_E)^2)d^4 k_E}{p_{0E}(k_E^2+m^2)(p_{0E}-2k_{0E})}
\ee
This integral has a pole at $k_{0E}=p_{0E}/2$. This results in the non-zero imaginary part,
\be 
I(p_E)=-\pi i\int\frac{(f(p_{0E}^2/4+\vk^2))^2 d^3 \vk}{p_{0E}(\vk^2+m^2+p_{0E}^2/4)}+P.V.\left(2\int\frac{f(k_E^2)\,f((p_E-k_E)^2)d^4 k_E}{p_{0E}(k_E^2+m^2)(p_{0E}-2k_{0E})} \right)
\ee
where $P.V.$ stands for the Principal Value definition of the integral.
Let us now continue analytically the function of $p_E$ to the physical momenta. Thus, we change $p_{0E}\rightarrow -i p_0$,
\be
\label{integral}
I(p)=-\pi \int\frac{(f(\vk^2-p_0^2/4))^2 d^3 \vk}{p_0(\vk^2+m^2-p_0^2/4-i\epsilon)}-2\int\frac{f(k^2)\,f((-ip_0-k_0)^2-\vk^2)d^4 k}{p_0(k^2+m^2)(p_0-2 i k_0)} 
\ee
This integral is real for $p_0^2<4 m^2$. In the opposite case, the first term would have an imaginary part, physically corresponding to the possibility of a particle production. The corresponding expression is,
\be 
{\rm Im}\,I(p)=-\pi^3 \frac{f(m^2)^2\sqrt{p_0^2-4m^2}}{p_0}\theta(p_0^2-4 m^2)
\ee
Then, taking into account that non-local function should be normalized in such a way that $f(m^2)=1$ to preserve the value of the residue of the propagator, we recover the familiar result for the local theory,
\be 
{\rm Im}\,{\cal M}=\frac{\lambda^2}{32\pi}\frac{\sqrt{p_0^2-4m^2}}{p_0}\theta(p_0^2-4 m^2).
\ee

{It is thus obvious that one observes the canonical optical theorem statement which tells us that the imaginary part of some Feynman graph is equivalent to the product of expressions corresponding to parts in which a given graph would be cut.}

We can demonstrate that the same result for the amplitude can be obtained if one take the formal definition of non-local amplitude in Minkowski space,
\be
\label{amp_M}
{\cal M}=-i\frac{\lambda^2}{32\pi^4}I(p)
\ee
Here 
\be 
I(p)=\int d^4 k \frac{f(k^2)\,f((p-k)^2)}{(k^2-m^2+i\epsilon)((p-k)^2-m^2+i\epsilon)}
\ee
and this integral over Minkowski momenta is divergent due to the inevitable presence of the essential singularity of function $f(\square)$ and consequently some growth direction at infinity as explained in the previous Section. However, applying formally the Wick rotation like in a local theory we arrive to the convergent integral. This integral can be written in the following form,
\be 
I(p)=-2\int\frac{f(k^2)\,f(({p}_M-k_E)^2)d^4 k_E}{p_0(k_E^2-m^2+i\epsilon)(k_0-p_0/2+i\epsilon)}
\ee
where we have not changed $k_0$ to $i k_{0E}$ in one instance to account the pole on the real axis of $k_0$ properly. Namely, we first have to transform
\be 
\frac{1}{k_0-p_0/2+i\epsilon}=-i\pi\delta(k_0-p_0/2)+P.V.\left(\frac{1}{k_0-p_0/2+i\epsilon}\right)
\ee
and then change in the second term $k_0\rightarrow ik_{0E}$. The other pole is not {pinched} by the contour. After that we arrive at
\be 
I(p)=\pi i\int \frac{d^3\vk (f(p_0^2/4-\vk^2))^2}{p_0(\vk^2+m^2-p_0^2/4-i\epsilon)}+2i\int\frac{f(k_E^2)\,f(({p}-k_E)^2)d^4 k_E}{p_0(p_0-2ik_{0E})(k_E^2+m^2)}
\ee 
which differs from \eqref{integral} by the factor $-i$ and this is precisely compensated by $i$ in the definition of amplitude \eqref{amp_M}.

Thus, the accurate definition of non-locals amplitudes through Euclidean integrals analytically continued to Minkowski external momenta is equivalent to the familiar one used for local theories. Even though the latter cannot be formalized through the Wick rotation in non-local theories because the Minkowski integral is divergent, this formal approach gives the same results. Further, as it was proven in \cite{Pius:2016jsl}, the way of amplitude's definition through Euclidean integrals preserves the unitarity of non-local theories and the optical theorem. Also one easily see that the local limit $f(\square)\to 1$ restores the standard textbook answers.
 
%%%%%%%%%%%%%%%%%%%%%%%%%%%%%%%%%%%%%%%%%%%%%%%%%%%%%%%%%%%%%%%%%%%%%%%%%%%%%%%%%%%%%
\section{Fish diagram 2-2 scattering}
\label{sec:simple}
In this section we consider the model with non-local function of the simplest form allowing
for the analytic computation of the amplitude, $f(\Box) = e^{\alpha\Box}$. The matrix element can be formally defined as
\be 
{\cal M}=-i\frac{\lambda^2}{32\pi^4}I(p)
\label{Mint}
\ee
where
\be 
\label{int}
I(p)=\int d^4 k\frac{e^{\alpha((p/2+k)^2+(p/2-k)^2)}}{((p/2+k)^2-m^2)((p/2-k)^2-m^2)}
\ee
As has been already discussed above, the Minkowski integral is divergent and the physical amplitude is defined
in another way. We replace loop momenta by their Euclidean counterparts as we would do for the Wick rotation in a local theory and then obtain $I(p)$ as a sum of the Euclidean integral and the pole part,
\be 
I(p)=I_{\cal E}(p)+I_{pole}(p)
\ee
The pole part here stands for the portion of the whole expression which corresponds to poles which appear at points $k_0=\pm p_0/2\pm\sqrt{\vec k^2+m^2}$ as long as these poles lie in the first and third quarter of the integration contour plane upon the prescription $m^2\to m^2-i\epsilon$.

Let us start with the computation of the Euclidean part,
{
\be 
\begin{split}
  I_{\cal E}=ie^{\alpha p^2/2}\int\frac{d k_0 d^3 k\,e^{-2\alpha (k_0^2+\vec k^2)}}{(k_0^2+\vec k^2+m^2-p^2/4+ i p k_0)(k_0^2+\vec k^2+m^2-p^2/4- i p k_0)}=\\
=2\pi i e^{\alpha p^2/2}\int \frac{d r d\theta r^3\sin^2{\theta} e^{-\alpha r^2}}{(r^2+m^2-p^2/4)^2+ p^2 r^2 \cos^2{\theta}}=\\
=\frac{4\pi^2 i e^{\alpha p^2/2}}{p^2}\int r dr e^{-2\alpha r^2}\l\sqrt{\frac{(r^2+m^2-p^2/4)^2+ p^2 r^2}{(r^2+m^2-p^2/4)}}-1\r
\end{split}
\ee
}
In the massless limit,
\be 
I_{\cal E}=\frac{4\pi^2 i e^{\alpha p^2/2}}{p^2}\l \int_0^{p^2/4-\epsilon_0}dz\,e^{-2\alpha z}\frac{z}{p^2/4-z} +\int_{p^2/4+\epsilon_0}^{\infty}dz\,e^{-2\alpha z}\frac{p^2/4}{z-p^2/4}\r
\ee
Evaluating these integrals one obtains the contribution to $\mathrm{Re}\,{\cal M}$
\be 
I_{\cal E}^{Re}=-\frac{2\pi^2 i}{ p^2 \alpha}\l e^{\alpha p^2/2}-1-2\alpha p^2 {\rm Ei}(\alpha p^2/2)+\alpha p^2(\gamma+\log{\epsilon_0})\r
\ee
Here $\gamma$ is Euler constant, $\mathrm{Ei}(z)$ is the integral exponent and we have taken the mean value of the integrals performing the limit $\epsilon_0\to 0$). As we will see later, the divergence $\log{\epsilon_0}$ will be exactly cancelled by the pole part of the amplitude in the massless limit.

\begin{figure}[htb]
    \begin{center}
      \includegraphics[scale=1.2]{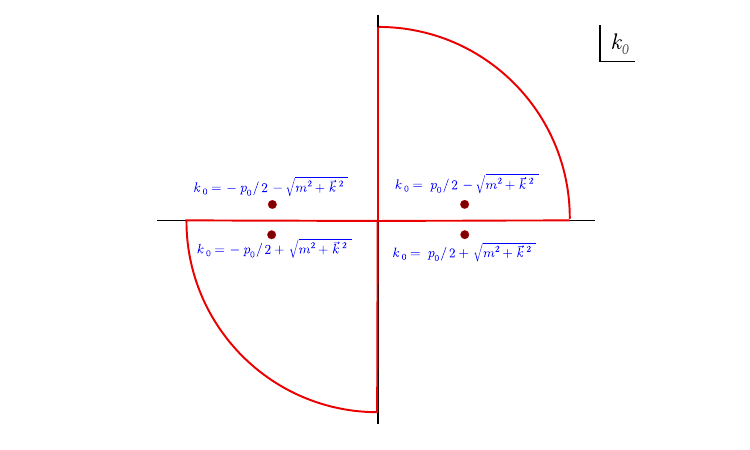}
  \caption{Poles of the amplitude (\ref{int}).}
\label{fig:poles}
\end{center}
\end{figure}

\begin{figure}[h!]
  \begin{center}
      \includegraphics[width=0.5\textwidth]{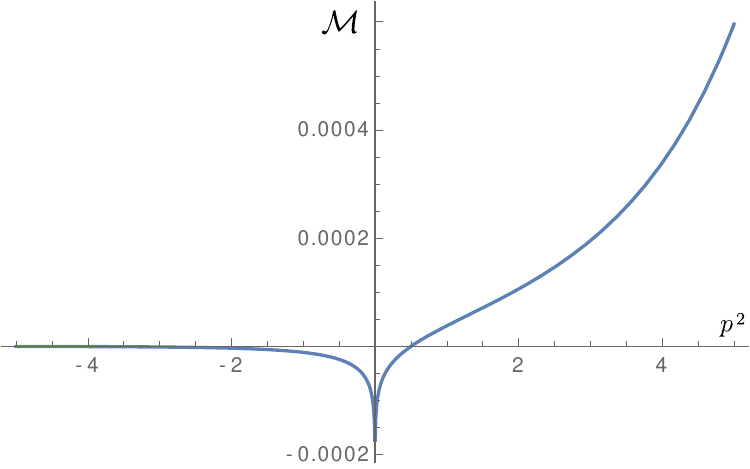}
  \end{center}
\caption{Amplitude (real part) dependence on the external momentum. Here $\alpha=1$, $m=0$, $\lambda=0.1$.}
  \label{fig:ampfish}
\end{figure}

The pole part is non-zero only when there are poles in I and III sectors of the complex plane,
i.e. when {$k^2 + m^2 > p^2/4$}. Fig.~\ref{fig:poles} illustrates this situation.
In this case, their impact to the amplitude is given by the sum of
the relevant residues in \eqref{int}
\be 
I_{pole}=-2\pi i \int d^3 k\l \Res(k_0=p_0/2-\sqrt{m^2+\vec k^2})-\Res(k_0=-p_0/2+\sqrt{m^2+\vec k^2})\r
\ee
which readily simplifies to
\be 
I_{pole}=-2\pi i\int\frac{d^3 k\,e^{2\alpha m^2+2\alpha p(p/2-\sqrt{\vec k^2+m^2})}}{2 p\sqrt{m^2+\vec k^2}(p/2-\sqrt{m^2+\vec k^2})}
\ee
For $m=0$ we obtain
\be 
I_{pole}=-\frac{8\pi^2 i e^{\alpha p^2}}{p^2}\int^{p^2-\epsilon_0}_{0}\frac{z dz e^{-4\alpha z}}{p^2/4-z}
\ee
We notice that integration up to $p^2$ follows from the fact that $k^2+m^2$ should be greater than $p^2$ in order to have poles in the I and III quarters of the integration plane.
\be 
I_{pole}= \frac{2\pi^2 i}{\alpha p^2}\l e^{\alpha p^2}-1-\alpha p^2{\rm Ei}(\alpha p^2)+\alpha p^2(\gamma+\log{\epsilon_0})\r
\ee
Summing up and adding the computed in Sec.~\ref{sec:unitarity} imaginary part contribution $-\pi^3$ to $I(p)$ we get the final answer\footnote{The rel part of this amplitude was also computed in \cite{Buoninfante:2018mre}}
\be 
I(p)=-\pi^3+\frac{ 2 i \pi^2}{\alpha p^2}\l e^{\alpha p^2}-\alpha p^2{\rm Ei}(\alpha p^2)-e^{\alpha p^2/2}+\frac{1}{2}\alpha p^2{\rm Ei}(\alpha p^2/2)\r
\ee
The real part of amplitude (\ref{Mint}) is plotted in Fig.~\ref{fig:ampfish} where presence of a strong coupling regime seems obvious. We will see in next Sections that this situations depends on the form-factor.

One can further observe using that for small arguments $\mathrm{Ei}(z)\approx \gamma+\log(z)+z$ that the local limit and the logarithmic singularity usually arising in the cut-off regularization restore as long as $\alpha\to 0$.

The total scattering amplitude including the crossed diagrams can be obtained as,
\be 
{\cal M}_{tot}=-i\frac{\lambda^2}{32\pi^4}\l I(\sqrt{s})+I(\sqrt{t})+I(\sqrt{u})\r
\ee 
where standard Mandelstam variables are used.

%%%%%%%%%%%%%%%%%%%%%%%%%%%%%%%%%%%%%%%%%%%%%%%%%%%%%%%%%%%%%%%%%
\section{Computation for a generic propagator}

The computation of the previous Section can be extended to a generic propagator, i.e. generic form-factor $f(\square)$. This can be done by utilizing the $x$-representation for propagators. This yields in our case the following quite a simple expression for the matrix element
\begin{equation}
  {\cal M}(p)=\frac{\lambda^2}{2(2\pi)^4}\int d^4x\Delta(x)^2e^{ipx}+i\frac{\lambda^2\pi}{32}+\frac{\lambda^2}{16p}\int_{-p/2}^{p/2}f(2pq)dq
  \label{MDx}
\end{equation}
where $\Delta(x)$ is the propagator in the $x$-representation and the whole formula is in a sense the Fourier transform of the product of the propagators. {In the first term both Fourier transform and integration over $x$ are performed in the Euclidean space. The last term corresponds to the pole contribution and it has such a simple form only for even form-factors $f(-k^2)=f(k^2)$. They are of a special interest because they would simultaneously decay along both real and imaginary directions in $k$.}

The propagator in $x$-space can be easily written as
\begin{equation}
  \Delta(x)=\frac{i}{16\pi^4}\int d^4k\frac{f(k^2)}{k^2}e^{ikx}
  \label{Dx}
\end{equation}
and in contrast with a local theory the latter integral can be straightforwardly made well defined at short distances, i.e. small $x$.
First, we go to spherical coordinates and simplify the last expression as follows
\begin{equation}
  \Delta(x)=\frac{i}{4\pi}\int_0^\infty dkf(k^2)\frac{J_1(kx)}x
  \label{Dx1}
\end{equation}
where $J_1(z)$ is the Bessel function.

Then we get by direct substitution
\begin{equation}
  {\cal M}(p)=-\frac{\lambda^2}{64\pi^3p}\int_0^\infty{J_1(px)J_1(kx)J_1(qx)}f(k^2)f(q^2)dkdqdx+i\frac{\lambda^2\pi}{32}+\frac{\lambda^2}{32p^2}\int_{-p^2}^{p^2}f(z)dz
  \label{MM}
\end{equation}
One can perform an integration over $x$ in the first integral in (\ref{MM}) analytically and the result is as follows \cite{gradshtein2007,https://doi.org/10.1112/plms/s2-40.1.37}:
\begin{equation*}
  \int_0^\infty{J_1(px)J_1(kx)J_1(qx)}dx=\left\{
    \begin{aligned}
  &\frac1{2\pi}\frac{\sqrt{(p^2-(k-q)^2)((k+q)^2-p^2)}}{pkq}\text{ for }|k-q|<p<k+q\\
  &0\text{ otherwise}
    \end{aligned}
  \right.
\end{equation*}
One further is left with an integration over $k$ and $q$ in the first integral in (\ref{MM}). The domain of integration can be a bit more intuitively rewritten as $\{-k+p<q<k+p,0<k<p\}\cup\{k-p<q<k+p,k>p\}$.

The last term in (\ref{MM}) demonstrates a not very much expected property. For large $p$ as long as the form-factor is an integrable function this term falls universally as $\sim 1/p^2$ irrespectively of a particular form-factor. We will see below that for a large class of suitable functions this makes the leading and apparently universal contribution to the amplitude. 

\begin{figure}[htb]
    \begin{center}
  \includegraphics[scale=.8]{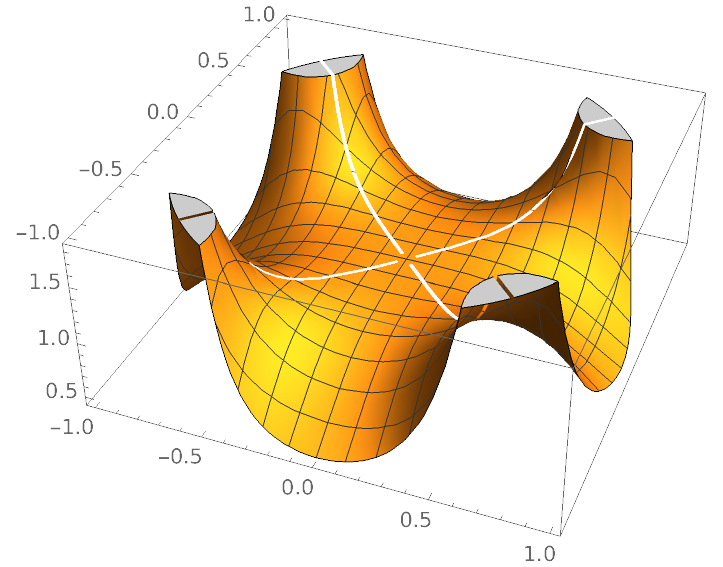}
\caption{Absolute value of the function $f_2({\rm Re}\,k,{\rm Im}\,k )$.}
\end{center}
\label{fig:mod}
\end{figure}

\begin{figure}[htb]
  \includegraphics[scale=.6]{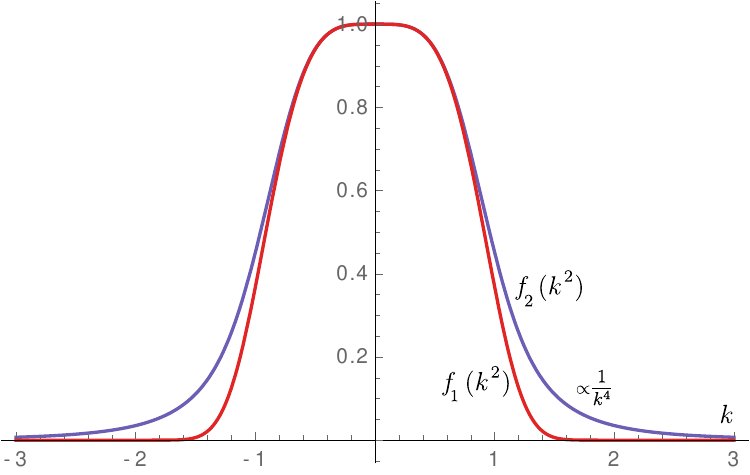}
\includegraphics[scale=.6]{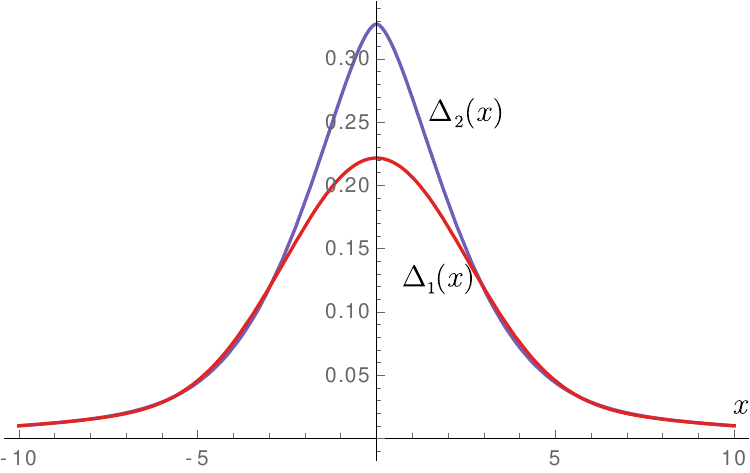}
\caption{Left plot represents the functions $f_1(k),~f_2(k)$ and right plot shows their Fourier images $\Delta_{1,2}(x)$.}
\label{fig:f}
\end{figure}

Proceeding by {computing the amplitudes} numerically for different form-factors $f(k^2)$ we choose the following two examples of non-local function. As the first example we take\footnote{Hereafter in our numerical studies we set the non-local scale to be unity.}
\begin{equation}
  f_1(k^2)=e^{-k^4}
\end{equation}
being the simplest non-local function allowing for staying in perturbative regime \cite{Chin:2018puw}. Besides that we consider a more involved function which falls polynomially along the real axis while it is an exponent of an entire function. This function was first suggested by Tomboulis in \cite{Tomboulis:1997gg} 

\be 
f_2(k^2)=e^{-\Gamma \left(0,k^4\right)-\gamma -\log{k^4}}
\ee 
Here $\Gamma(0,k^4)$ is an incomplete gamma function.
The latter form-factor having a polynomial decay at infinity making the theory avoiding a possible strong coupling regime. For large real $k$ one has $f_2(k^2)\propto 1/k^4$. In Fig.~\ref{fig:mod} we plot the absolute value of this function on the whole complex plane to make its growth directions visible.
Also we plot the Euclidean four-dimensional Fourier images $\Delta_1(x)$ and $\Delta_2(x)$ for the functions $f_1(k)$ and $f_2(k)$, respectively, in Fig.~\ref{fig:f}.

\begin{figure}[h]
\begin{center}
    \includegraphics[scale=.8]{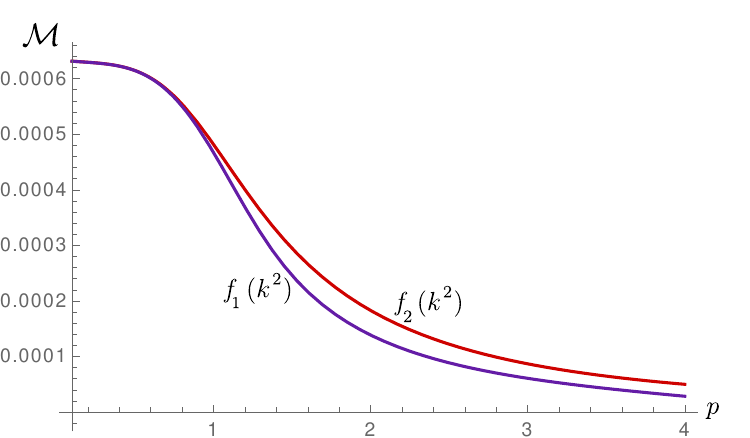}
    \end{center}
\caption{Scattering 2-2 one-loop amplitude as a function of the external momentum. Red curve is for form-factor $f_1(k)$ and blue curve is for form-factor $f_2(k)$. Here we have taken $\lambda=0.1$.}
\label{fig:amp}
\end{figure}

The resulting amplitude (real part) numerically computed with the use of (\ref{MM}) is plotted in Fig.~\ref{fig:amp}. One can see that the curves  fall for momenta much higher than non-locality scale\footnote{{The curves also grow for small momenta reproducing the logarithmic divergence in the local limit but this grows is not apparent for the chosen domain of the numeric integration.}}. However, as was mentioned above, the asymptotes of this falling are of the same order $\sim 1/p^2$ for both functions and are determined by the last term in expression (\ref{MM}). This term arises due to the pole structure in Minkowskian amplitudes. Further one can deduce numerically that the behavior of the first term in (\ref{MM}) depends on the UV behaviour of non-local propagator and differs for the two cases under consideration. For function $f_1$  it is an exponential fall as $\sim e^{-p^4}$ while for function $f_2$ it is the power law suppression $\sim p^{-6}$.
We also see that the model remains weakly coupled for both chosen form-factors in contrast with the $e^{\alpha p^2}$
form-factor discussed in the previous Section.

\section{UV properties of the amplitudes in non-local theory}

The UV behaviour of the scattering amplitudes is determined by the form of non-local propagator. As we obtained in Section~\ref{sec:simple}, for the propagator $e^{\Box/\Lambda^2}/\Box$ the amplitude is growing exponentially for positive $p^2$. Does it mean that this kind of model are not valid from the point of view of quantum field theory \cite{Tokuda:2019nqb}? We cannot immediately conclude that such kind of propagators cannot be considered because around the non-locality scale the model undergoes strong coupling regime which means that one-loop approximation is not valid anymore. However, this does not mean that the unitarity is broken as long as non-perturbative methods can in principle lead to a result which is still consistent with unitarity. We leave this question for future study.

In this work we concentrate on the results which can be obtained perturbatively. If the propagator $f(k^2)/k^2$ is falling for both signs of $k^2$ than we expect the loop contributions to fall for all physical values of external momenta. However, the scattering amplitude would inevitably grow for some directions of $s$ and $t$ in the complex plane. Let us check the consistency of such a behaviour with the known results about the properties of the amplitudes. For example, the Martin-Froissart bound \cite{Martin:1965jj} tells us that the amplitude in the forward limit $t\rightarrow 0$ is bounded by
\be
|{\cal M}(s,0)|< C s (\log s)^2
\ee
on the $s$ complex plane. Although this bound was obtained in local theories, the recent work \cite{Tokuda:2019nqb} provides a formal proof of the similar bound in non-local theories,
\be 
\label{bound}
|{\cal M}(s,0)|\lesssim C s^{1+2a}.
\ee
This bound was obtained within rather weak assumptions including the validity of partial wave expansion, unitarity, analiticity and the exponential boundedness $|{\cal M}(s,t)|<s^{N}e^{a \,|s|^{a}}$ for fixed $t$. Here $\alpha, ~N,~C$ are some positive constants. These assumptions except the last one are expected to be fulfilled in the models we consider in this work and in more generic settings like in \cite{Pius:2016jsl}. The bound \eqref{bound} actually means that the amplitudes in the forward limit are {\it polynomially} bounded while one would expect the exponential growth in some directions on the complex $s$ plane. 

Let us discuss this issue using the simple model with cubic interaction where the exponential behaviour shows up already at tree level.
\be 
\label{phi3}
S=\int d^4x \left[{\frac{1}{2}\phi (\square-m^2)f(\Box)^{-1}\phi-\frac{\lambda}{3!}\phi^3}\right]
\ee
The tree level amplitude in this model has the form
\be 
{\cal M}(s,t)\sim \lambda^2\left(\frac{f(s)}{s-m^2}+\frac{f(t)}{t-m^2}+\frac{f(u)}{u-m^2}\right)
\ee
Thus, $f(s)$ must have a direction of exponential growth since an entire function which is polynomially bounded on the whole complex plane is nothing else than polynomial of finite order. However, for all physical momenta the amplitude is small if we require $f(q^2)$ to fall for real $q^2$. Thus, with this choice, the model stays in perturbative regime for physical values of momenta. However, if one computes the amplitude for complex values of momenta this could lead to breaking of the perturbation theory. The naive computation leading to the exponential growth can be invalid anymore. Therefore the conclusion about the amplitude's exponential growth in the nonphysical region is not correct.
%Moreover, the analyticity arguments leading eventually to the bound \eqref{bound} allows us to conclude that the full nonperturbative amplitude is bounded polynomially. 

The bound \eqref{bound} leads to another interesting consequences for the model \eqref{phi3}. Under the condition of polynomial boundedness of the amplitude in the forward limit we have the Cerulus-Martin lower bound on the amplitude falling \cite{Cerulus:1964cjb},
\be 
\label{cerulus}
|{\cal M}(s,t\rightarrow 0)|>C e^{-\gamma \sqrt{s}\log (s/s_0)}
\ee 
Here $C,~\gamma,~s_0$ are positive constants. This bound is obviously violated if we take $f(k^2)\propto e^{-\alpha k^4}$. The tree level amplitude is falling too fast while the model is in a perturbative regime so we can trust this computation for physical momenta. This contradiction means that something is wrong with the model with an exponentially falling propagator.
%We leave for the future study finding which assumption is actually broken in a model with $f(k^2)\propto e^{-\alpha k^4}$.
Notice that in the work \cite{Tomboulis:1997gg} the choice of polynomially suppressed propagators was suggested motivated by different arguments such as of power counting for example. The propagator falling as $1/k^n$ is still consistent with the discussed bounds. 

Notice that disfavouring of the exponentially suppressed propagator is not applicable for models with contact $\phi^4$ interactions since the leading term is given by a constant tree level contribution. This kind of suppression for those models is not in a contradiction with Martin-Cerulus bound \cite{Cerulus:1964cjb}.

\section{Conclusion and Outlook}

We have shown explicitly that the optical theorem statement holds in non-local higher derivative theories upon utilizing a modified computation prescription. The prescription consist of computing Feynman graphs assuming all the internal momenta are Euclidean and then continuing the final answer analytically to Minkowski signature for external momenta. Resulting expressions satisfy the optical theorem statement and obey the correct local theory limit. We were able to show that the one-loop correction to the vertex in a non-local $\phi^4$ theory can be computed for arbitrary form-factors and have presented both analytic and numeric derivations of the result. {We found that, independently of the non-local form-factor, the leading term in the amplitude at high energies falls as $\sim 1/p^2$ as long as $f(k^2)=f(-k^2)$. This term arises as a result of computation of the pole contribution to the amplitude. Thus, non-local amplitudes for even form-factors at one loop have a universal behaviour in the UV limit. }

As an important observation we have found that under an assumption that the Martin-Cerulus bound \cite{Cerulus:1964cjb} are satisfied in non-local models as in local ones, than the amplitudes should not fall too fast obeying bound (\ref{cerulus}). If such a non-local theory has a contact interaction than this bound can always be satisfied irrespectively of the decay rate of the propagator as the contact interaction term always dominates in the perturbative expansion. On the other hand, in the absence of the contact interaction (for instance computing corrections to 2-2 scattering in $\phi^3$ theory) one must demand that propagator decays polynomially favoring this way the type of functions suggested in \cite{Tomboulis:1997gg}. 

{As the next step one would aim at demonstrating that the background field method provides consistent results and thus can be applied to AID models which is a plausible expectation. Also the strong coupling regime which may arise in such non-local field theories \cite{Koshelev:2020fok} due to the growth of the amplitudes for complex parameters is an important question for further investigations. Yet another ambitious problem is to study such theories in the large-$N$ approximations \cite{tHooft:2002ufq} and see how they are compared to their local counterparts.}

\section*{Note added}

Upon completion of the manuscript we have met with great interest a related study arXiv:2103.00353 discussing the same problem of unitarity of non-local field theories.

\section*{Acknowledgements}

Authors would like to thank I.~Aref'eva and M.~Shaposhnikov for comments and questions on the manuscript.
AK is supported by FCT Portugal investigator project IF/01607/2015. AT is supported by the Academy of Finland grant 318319. The part of the work performed by AT related to the numeric computations of the amplitudes is supported by the Russian Science Foundation grant 19-12-00393.

%\bibliographystyle{utphys}
%\bibliography{anya}

\providecommand{\href}[2]{#2}\begingroup\raggedright\endgroup

\end{document}